\documentclass[]{sn-jnl}

\usepackage{amsmath}
\usepackage{color}
\usepackage{float}
\usepackage{bm}
\usepackage{hyperref}
\usepackage{soul}
\usepackage{xcolor}

\usepackage{geometry}
\usepackage[style=nature,
            ]{biblatex}
            
\usepackage{amssymb}

\title[Electromagnetic radiation]{Electromagnetic radiation by turbulent, magnetized and
randomly inhomogeneous solar radio sources generated by electron beams}

\begin{document}

\author*[1,2]{C. Krafft}
\email{catherine.krafft@universite-paris-saclay.fr}

\author[1]{A. S. Volokitin}

\author[1]{F. J. Polanco-Rodr{\'{i}}guez}

\author[1]{P. Savoini}

\affil[1]{Laboratoire de Physique des Plasmas (LPP), CNRS, Sorbonne Université, Observatoire de Paris, Université Paris-Saclay, Ecole polytechnique, Institut Polytechnique de Paris, 91120 Palaiseau, France}
\affil[2]{Institut Universitaire de France (IUF)}

\abstract{During Type III solar radio bursts, electromagnetic waves are radiated at plasma frequency $\omega_p$ and its harmonics by electrostatic wave turbulence generated by electron beams ejected by Sun in randomly inhomogeneous solar wind and coronal plasmas. These emissions, detected since decades by spacecraft and radiotelescopes, are split by the plasma magnetic field into three modes $\mathcal{X}$, $\mathcal{O}$ and $\mathcal{Z}$ of different dispersion, polarization and radiation properties. This work demonstrates, using three independent and converging approaches, that only a small fraction of electromagnetic energy radiated at $\omega_p$ ($\lesssim10\%$) is escaping from beam-generated radio sources, mainly as $\mathcal{O}$-mode waves and, depending on plasma conditions, as $\mathcal{X}$-mode waves. Most energy is radiated in  $\mathcal{Z}$-mode and can therefore be only observed close to sources. Results have major implications for solar radio emission and provide strong support for interpretation of observations performed up to close distances to Sun by Parker Solar Probe and Solar Orbiter spacecraft.}

\maketitle

\section{Introduction}
During Type III solar bursts, which are among the most intense radio sources in
 heliosphere \cite{Dulk1985,ReidRatcliffe2014,Melrose2017}, energetic electron beams are accelerated 
in the solar atmosphere, propagate along open magnetic field lines in the interplanetary
space, generate electrostatic upper-hybrid wave turbulence which in turn
radiates electromagnetic waves at the plasma frequency $\omega_{p}$ and
its harmonics $n\omega_{p}$. Such radio emissions are routinely observed (e.g.  \cite{ReidKontar2021,Chen2021,Badman2022,Jebaraj2023a,Lorfing2023,Krupar2024a,Dabrowski2024,Vecchio2024}, and references therein) by space-born and ground-based instruments as, to cite only recent and exceptional
ones, the satellites Parker Solar Probe and Solar Orbiter \cite{Fox2016,Muller2020}, as well as
the radiotelescope LOFAR and its extension NenuFAR \cite{VanHaarlem2013}. The aim of this work is to provide decisive answers to open questions about the properties of these electromagnetic emissions and their generation mechanisms.

The solar wind and corona are randomly inhomogeneous plasmas where density
fluctuations $\delta n$ of average levels $\Delta N=<(\delta n/n_0)^2>^{1/2}$ up to several percent of the
ambient plasma density  $n_0$ have been measured \cite{Celnikier1987,Krupar2018,Krupar2020}. Their impact on beam-plasma interactions
and subsequent processes responsible for radio emissions has been shown to
be significant, and even crucial for fundamental electromagnetic radiation
at $\omega _{p}$ 
(e.g. \cite{Kontar2001,Krafft2013,ReidKontar2013,VoshchepynetsKrasnoselskikh2015,VolokitinKrafft2018,Krasnoselskikh2019,KrafftSavoini2022a}). Several mechanisms for the generation of radio emission 
by Type III bursts have been proposed and discussed over past decades. 
To date, nonlinear wave-wave interactions, processes of mode
conversion from electrostatic into electromagnetic waves or so-called antenna mechanisms 
have been shown to describe various features of
observed electromagnetic emissions at different solar wind
conditions. Three-wave interactions as decay or coalescence of
waves (e.g. \cite{Melrose2017,Tsytovich1970,KrafftSavoini2021,KrafftSavoini2024,Krafft2024}) are generally dominant in
quasi-homogeneous plasmas with $\Delta N\lesssim 3(v_{T}/v_{b})^{2}$ \cite{Krafft2013,Ryutov1970}, where $v_{T}$ and $v_{b}$ are the electron
plasma thermal and beam drift velocities, respectively. In randomly
inhomogeneous plasmas with $\Delta N\gtrsim
3(v_{T}/v_{b})^{2},$  wave transformation processes as mode conversion,
refraction, reflection, tunneling, and scattering of electrostatic wave
turbulence on density fluctuations are thought to be dominant (e.g. \cite{VolokitinKrafft2018,Krasnoselskikh2019,KrafftSavoini2022a,Field1956,VolokitinKrafft2020}).

Conversion processes between wave modes play an important role in space and
astrophysical plasmas (e.g. \cite{GurnettKurth1996,LaBelleTreumann2002}). In
particular, during linear mode conversion (LMC), electrostatic wave energy
can be transformed at constant frequency into electromagnetic energy when interacting with
density gradients or via scattering on random density fluctuations.
Such process is relevant to Type II and III radio bursts in the solar corona
and wind, as well as to ionospheric and Earth or planetary magnetospheric
physics (e.g. \cite{LaBelleTreumann2002}).

In particular, and this is the case of interest here, electrostatic upper-hybrid wavepackets  (also known as Langmuir/$\mathcal{Z}$-mode or
Langmuir/slow-extraordinary waves, referred to below as $\mathcal{LZ}$
waves) excited
by beam-plasma interactions during Type  III radio bursts can be transformed near $\omega _{p}$ into ordinary as well as fast
and slow extraordinary electromagnetic modes $\mathcal{O}$, $\mathcal{X}$
and $\mathcal{Z}$, respectively, depending on plasma and beam characteristics (the two latter modes do not exist in unmagnetized plasmas). These three modes present different radiation, dispersion and polarization properties. It is worth noting that $\mathcal{LZ}$ waves, with
frequencies $\omega \gtrsim \omega _{p}$, are commonly observed in solar
wind and Earth foreshock plasmas, where their role and properties are actively
debated, especially their transition into electromagnetic $\mathcal{Z}$-mode waves at frequencies $\omega \lesssim \omega_p$ (e.g. \cite{Bale1998,Malaspina2011,Kellogg2013,GrahamCairns2013}).
However, the first experimental identification of $\mathcal{Z}$-mode waves' magnetic component has only been achieved very recently \cite{Larosa2022}. On the other
hand, electromagnetic emissions at $\omega_{p}$ from Type III solar bursts have
been observed in the form of $\mathcal{X}$- and $\mathcal{O}$-mode waves,
which are able to escape from radio sources where they are generated
(e.g. \cite{Jebaraj2023a,Pulupa2020}). However, due to their
dispersive properties, $\mathcal{Z}$-mode waves remain mostly within or close to their generation regions.

Theoretical and numerical works have been performed to study electromagnetic
emissions via LMC, by calculating the radiation resulting from the
interaction of incoming monochromatic electrostatic waves with a density gradient, for
different plasma magnetic fields, gradients' characteristics, angles of
incidence of waves,...
\cite{Oya1971,Hinkel-Lipsker1989,Yin1998,Kim2008,Kim2013,Schleyer2013,Schleyer2014}. All of them concluded that electromagnetic radiation occurs
in the form of $\mathcal{X}$- and $\mathcal{O}$-mode waves. However, no
studies on the radiation at $\omega_p$ of electrostatic wave turbulence generated by beam-plasma instability in magnetized and
randomly inhomogeneous plasmas have been carried out to date. This paper fills this gap for weakly
and moderately magnetized plasmas.

Using three independent methods
(Particle-In-Cell simulations, theoretical and numerical model, as well
as analytical calculations performed in the framework of weak turbulence theory
extended to randomly inhomogeneous plasmas), we came to new and converging
conclusions. Electromagnetic waves in $\mathcal{Z}$-mode are generated in the radio
sources with the highest rate and saturation energy, which are at least
an order of magnitude higher than in $\mathcal{O}$-mode. $\mathcal{X}$-mode waves are only emitted in plasmas where
the ratio of cyclotron to plasma frequencies $\omega _{c}/\omega _{p}$ is
lower than the average level of density fluctuations $\Delta N$; this
condition strongly limits  their appearance during Type III solar radio
bursts. Then, only a small fraction of energy radiated at $\omega_p$ ($\lesssim10\%$) is escaping from beam-generated radio sources, mainly in the form of $\mathcal{O}$-mode waves and, much more rarely, of $\mathcal{X}$-mode ones, whereas the main part remains inside or close to sources. These findings have major implications for
solar radio emission, and provide support for interpretation of observations performed at close distances to Sun by satellites as Parker Solar Probe and Solar Orbiter.

\section{Results}
Large-scale and long-term high-performance 2D/3V Particle-In-Cell
simulations of high wavenumber and frequency resolutions are performed to
evidence electromagnetic emissions by upper-hybrid wave turbulence generated
by an electron beam in randomly inhomogeneous and weakly to moderately
magnetized plasmas. Physical conditions used are relevant to
Type III radio bursts in the solar wind with $0.02\leq \omega
_{c}/\omega _{p}\leq 0.5$, corresponding to a wide range of distances to the Sun ($\sim0.05-1$ AU). The plasma is initially set 
inhomogeneous, with  random density fluctuations of average levels $0<\Delta
N\lesssim 0.05$ and wavelengths much larger than those of $\mathcal{LZ}$ waves \cite{Celnikier1987}. More details are given below in the section "Methods".

\begin{figure}
    \centering
    \includegraphics[width=1\linewidth]{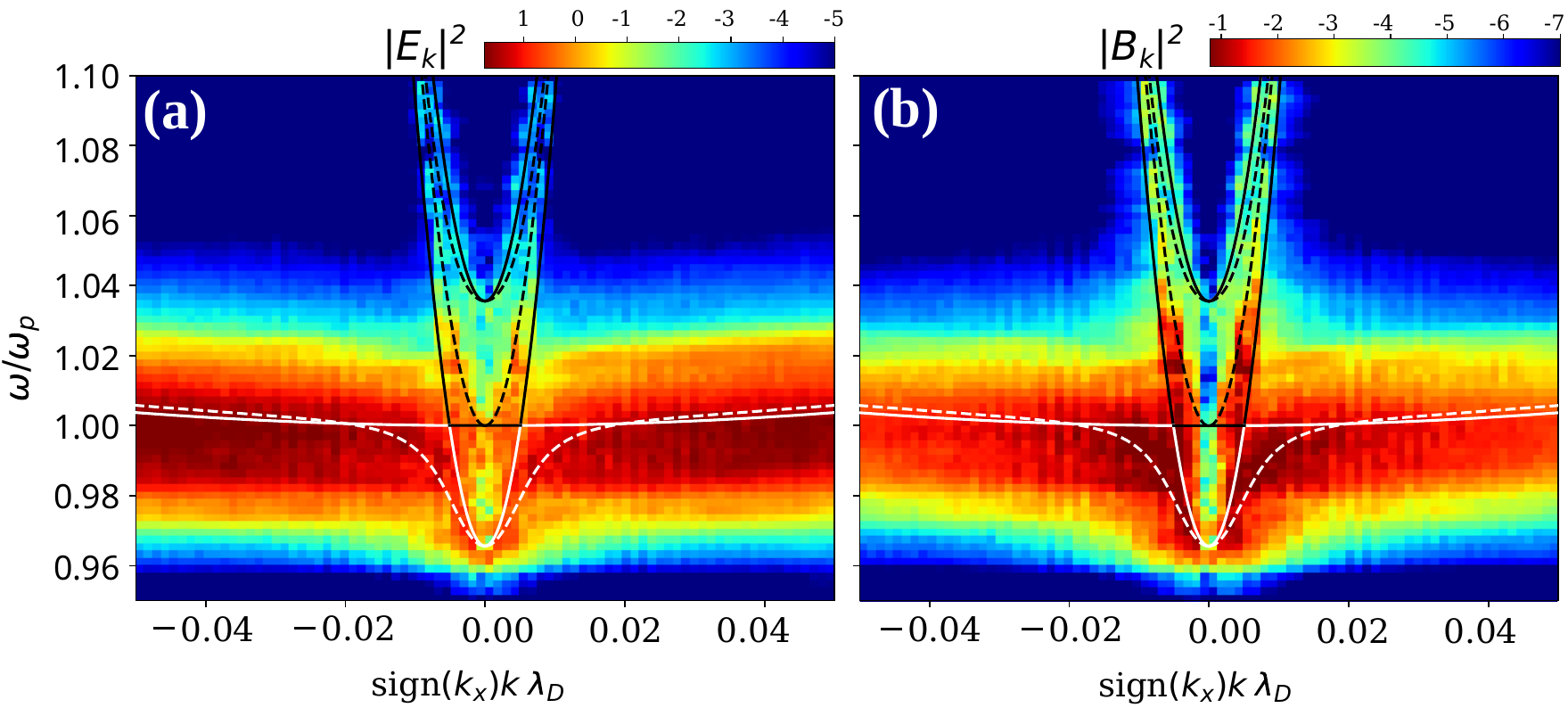}
    \caption{Dispersion of the spectral wave energy (integrated on the
propagation angle $\theta $ between $\mathbf{k}$ and the ambient magnetic
field $\mathbf{B}_0$) as a function of the normalized wave frequency $%
\omega /\omega _{p}$ and wavenumber modulus $sign(k_{x})k\lambda _{D},$ at
time $\omega _{p}t\simeq 5000$, for $\Delta N=0.025$ and$\ \omega
_{c}/\omega _{p}=0.07;$ $k_{x}$ is the wavenumber along  $\mathbf{B}_0$. (a) : Spectral electric energy $\left\vert \mathbf{E}_\mathbf{k}\right\vert
^{2}$. (b) : Spectral magnetic energy $\left\vert \mathbf{B}_\mathbf{k}\right\vert
^{2}$. Both energies are presented in arbitrary units and logarithmic scales.
Superimposed dispersion curves of the $\mathcal{O}$-, $\mathcal{X}$- and $%
\mathcal{Z}$-modes are calculated using the magnetoionic theory (black lines for $\mathcal{O}$- and $\mathcal{X}$-mode waves, and white lines for $\mathcal{Z}$-mode and $\mathcal{LZ}$ waves, respectively) and represented for the parallel ($\theta =0$) and
perpendicular ($\theta =\pi /2$) propagation angles by solid and dashed
lines, respectively. Numerical simulation parameters are : $%
L_{x}\times L_{y}=5792\times 2896\lambda _{D}^{2}$ (box lengths)$,$ $\delta
k_{\parallel }=0.001\lambda _{D}^{-1},$ $\delta k_{\perp }=0.002\lambda
_{D}^{-1}$ and $\delta \omega =0.002\omega _{p}$ (wavenumbers' and frequency
resolutions).}
    \label{fig1}
\end{figure}

Fig. \ref{fig1}a shows the dispersion of the spectral electric energy $%
\left\vert \mathbf{E}_\mathbf{k}\right\vert ^{2}$ as a function of wave
frequency $\omega $ and wavenumber $k=|\mathbf{k} |,$ at time $\omega _{p}t\simeq
5000$, for a simulation with $\Delta N=0.025$ and $\ \omega _{c}/\omega _{p}=0.07$;
 dispersion curves of $\mathcal{O}$-, $\mathcal{X}$- and $%
\mathcal{Z}$-modes, calculated in a homogeneous plasma for the parallel  ($\theta=0$)
and perpendicular ($\theta=\pi/2$)
propagation angles, are superimposed (solid and dashed lines, respectively).
One observes that electrostatic $\mathcal{LZ}$ waves are excited by the beam along their
dispersion curve (at $k\lambda_D \gtrsim 0.02$), acquiring more and more magnetic energy with decreasing frequency (see Fig. \ref{fig1}b showing the dispersion of the  spectral magnetic energy $\left\vert \mathbf{B}_\mathbf{k}\right\vert ^{2}$), until they become electromagnetic 
$\mathcal{Z}$-mode waves with frequencies lying within the range $\omega _{c\mathcal{Z}}\lesssim
\omega \lesssim \omega _{p}$ ($\omega _{c\mathcal{Z}}\simeq \omega
_{p}-\omega _{c}/2\simeq 0.975$ is the $\mathcal{Z}$-mode\ cutoff
frequency). Note however that for larger plasma magnetic fields $\omega_c/\omega_p \gtrsim 0.14$, most of $\mathcal{Z}$-mode emission does not reach $\omega _{c\mathcal{Z}}$.  The observed frequency broadening $%
\Delta \omega \simeq \omega _{p}\Delta N$ at $k\lambda_D \gtrsim 0.02$ is due partially to $\mathcal{LZ}$ waves' scattering on random density fluctuations;
their spectral energy $%
\left\vert \mathbf{E}_\mathbf{k}\right\vert ^{2}$ is shifted down to lower frequencies (compare with theoretical dispersion
curves) as it accumulates over time within plasma density depletions \cite{krafftVolokitin2021}. 
Linear mode conversion (LMC) 
at constant frequency around $\omega_{p}$ is responsible for $\mathcal{LZ}$
waves' transformations into electromagnetic radiation mainly carried by oblique $%
\mathcal{O}$- and $\mathcal{Z}$-mode waves. The former are excited up to
frequencies $\omega \simeq 1.04\omega _{p}$ (Figs. \ref{fig1}a,b), and can also be produced 
- with significantly lower growth rates - via
electromagnetic decay (nonlinear three-wave interaction), which can be
stimulated by LMC \cite{Krafft2024}. Finally, note the absence of $%
\mathcal{X}$-mode radiation above its cutoff frequency $\omega \geq \omega
_{c\mathcal{X}}\simeq \omega _{p}+\omega _{c}/2\simeq 1.035\omega_p.$

\begin{figure}[H]
    \centering
    \includegraphics[width=\linewidth]{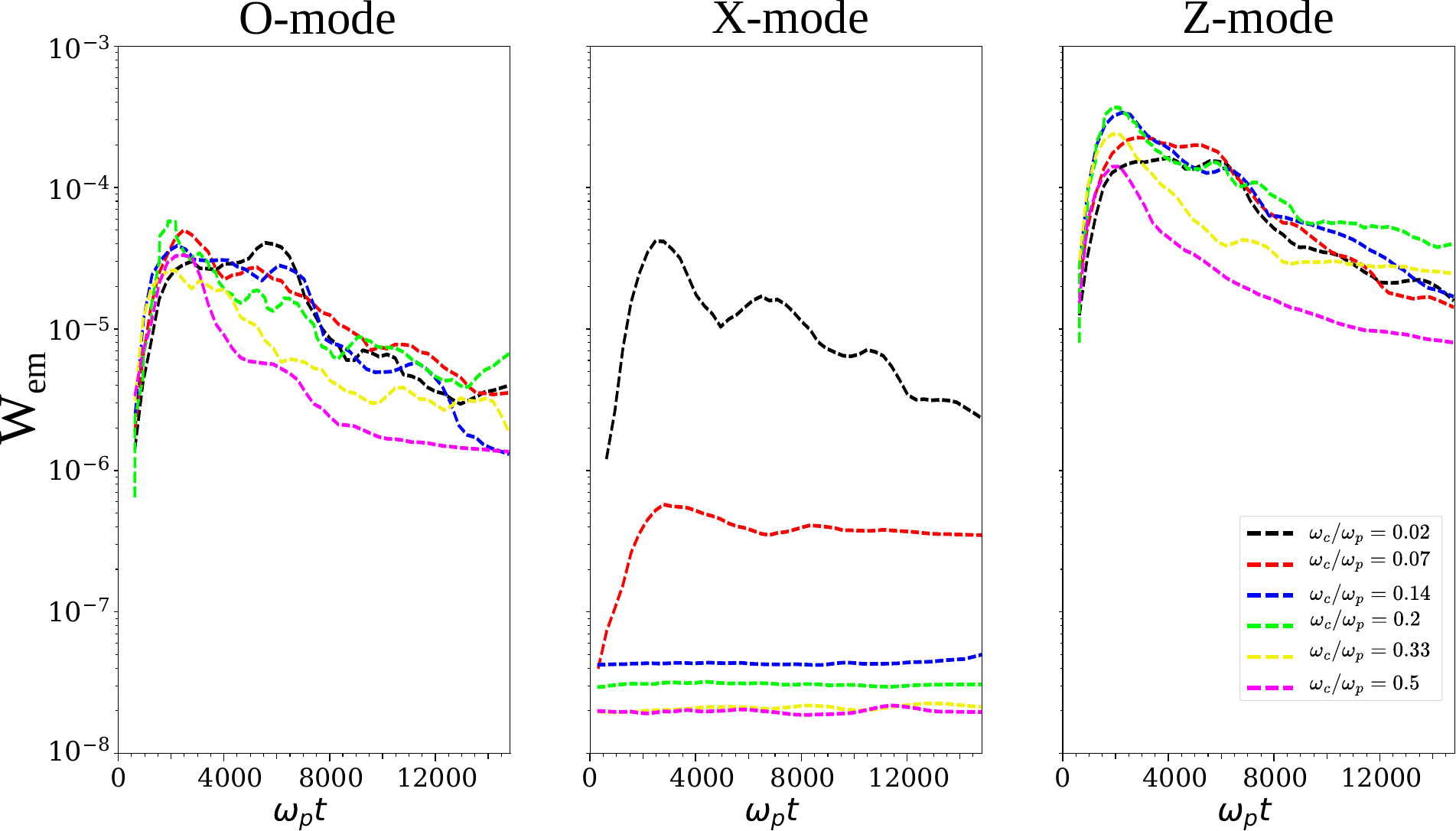}
    \caption{Variations with the normalized time $\omega _{p}t$ of the electromagnetic energy $W_{em}$ radiated around frequency $\omega _{p}$ in $\mathcal{O}$-, $\mathcal{X}$- and $\mathcal{Z}$-modes (panels (a), (b), and (c), respectively), in logarithmic scales, integrated on the 2D simulation box and normalized by the initial beam energy, in plasmas with $\Delta N=0.05$ and different ambient magnetic fields, i.e. $\omega _{c}/\omega _{p}=0,$ $0.02$, $0.07,$ $0.14,$ $0.2$, $0.33,$ and $0.5$ (see legend in (c)). The three panels (a)-(c) have the same vertical and horizontal scales. The energies $W_{em}$ are calculated by applying appropriate filtering to wave dispersion and spectra.}
      \label{fig2}
\end{figure}

Figs. \ref{fig2}a-c show the variations with time of the electromagnetic
energy $W_{em}(t)$ radiated around $\omega_{p}$ in $\mathcal{O}$-, $\mathcal{X}$- and $\mathcal{Z}$%
-modes, normalized by the initial beam energy, for $\Delta
N=0.05$ and different plasma magnetic fields $\omega _{c}/\omega _{p}=0,$ $0.02$, $%
0.07,$ $0.14,$ $0.2$, $0.33$, and $0.5$. Note that $\mathcal{Z}
$-mode energy includes contributions of these waves near $\omega_p$, i.e. within the frequency range $\omega_m\lesssim\omega < \omega _{p}$ ($\omega_m=\max(\omega_{c\mathcal{Z}}-\Delta\omega, 0.9)$) and the wavenumber interval $|k| \lesssim k_{\ast \perp } \simeq 0.02$, where $k_{\ast \perp }$ is the
perpendicular wavenumber modulus at $\omega =\omega _{p}$. 

Several important statements can be formulated at this stage. First, $\mathcal{Z}$-mode waves exhibit the highest 
growth rates and saturated energies. For $0.02\leqslant \omega _{c}/\omega _{p}\leqslant 0.2$, their 
 energies present close behavior and values, but decrease with increasing $\omega _{c}/\omega _{p}$ for $\omega
_{c}/\omega _{p}> 0.2$. After saturation, all waves lose energy due partly to
damping of upper-hybrid wave turbulence, which results from $\mathcal{LZ}$
wave scattering on density fluctuations leading to the formation of a tail of
accelerated beam electrons \cite{KrafftSavoini2023}. Second, energies carried by $\mathcal{O}$-mode waves are approximately an order of magnitude lower than those of $\mathcal{Z}$-mode waves, and their
growth rates are also smaller (as confirmed also by Figs. \ref{fig3}-\ref{fig5} below); 
they weakly depend on magnetization. Third,
energies of $\mathcal{X}$-mode waves reach significant levels,
between those of $\mathcal{O}$- and $\mathcal{Z}$-mode waves, only for 
$\omega _{c}/\omega _{p}=0.02$, and are negligibly small for other values of $\omega _{c}/\omega _{p}$; at $\omega _{c}/\omega _{p}=0.07,$  $\mathcal{X}$-mode
energy does not reach the noise level but is very small. Note that we find and discuss below a relation between $\omega _{c}/\omega _{p}$ and $\Delta N$ presenting the conditions for suppression or appearance of $\mathcal{X}$-mode radiation.

At this stage we have shown the prominence of energy carried by $\mathcal{Z}$-mode waves with respect to other modes as well as the possible suppression of $\mathcal{X}$-mode radiation due to combined effects of plasma magnetization and random density inhomogeneities. In order to confirm and complete by a second and independent approach the above results provided by selfconsistent PIC simulations, we have built a new theoretical and numerical model.
Let us consider a magnetized plasma source with external random
density fluctuations and study the dynamics of an initial spectrum of electrostatic upper-hybrid wave turbulence (which can be anisotropic to mimic wave excitation by an electron beam) using modified Zakharov equations involving
weak magnetic effects, i.e. with $\omega _{c}/\omega _{p}\lesssim 0.2$ \cite{KrasnoselskikhSotnikov1977}. When density fluctuations of finite average level $\Delta N$ are present in the plasma, three-wave nonlinear
interactions are not the dominant processes of electromagnetic wave generation at $\omega _{p}$ \cite{KrafftSavoini2024,Krafft2024}. Indeed high-frequency upper-hybrid waves are mainly involved in their interactions with the slowly varying density
fluctuations, i.e. in processes of wave transformations as
mode conversion, angular scattering, reflection, refraction, tunneling, and
trapping in density depletions. Electromagnetic waves are radiated within the randomly inhomogeneous plasma source at frequency $\omega _{p}$ via linear mode conversion at constant frequency (LMC) and are further leaving it to propagate away through a uniform plasma, without interacting back with the density fluctuations.

This model provides a compact equation describing the time evolution in $\mathbf{k}$-space of the slowly varying envelope $B_{\mathbf{k}}^{\pm }(t)$ of the magnetic field $\mathbf{B}(\mathbf{r},t)$ radiated in the $\mathcal{X}$- and $\mathcal{Z}$-mods  (see the Section "Methods" for details)
\begin{equation}
\left( i\frac{\partial }{\partial t}-\Delta \omega _{\pm }\right) B_{\mathbf{%
k}}^{\pm }\simeq-\frac{2\pi c}{\omega _{p}}\mathbf{a}_{\mathbf{k}}^{\pm\ast}\cdot 
\mathbf{k}\times \left( \delta \mathbf{j}_{\mathbf{k}}\pm i\mathbf{h}\times
\delta \mathbf{j}_{\mathbf{k}}\right) ,  \label{maineq}
\end{equation}%
where the superscript '$\ast $' indicates complex conjugation; $\delta 
\mathbf{j}_{\mathbf{k}}$ is the Fourier component of the current generated in the radio source as a result of
$\mathcal{LZ}$ waves' scattering on density fluctuations, $\mathbf{h}=\mathbf{B_0}/B_0$ is the plasma magnetic field direction, $\Delta \omega_{\pm }=\omega _{\mathbf{k}}^{\pm }-\omega_p$  is the frequency detuning and $\omega _{\mathbf{k}}^{\pm }$ is the dispersion relation of $\mathcal{X}$- and $\mathcal{Z}
$-mode waves near their cutoff frequencies; $\mathbf{a}_\mathbf{k}^{\pm }$ is the polarization vector; the signs $+$ and $-$ correspond to the $\mathcal{X}$- and  $\mathcal{Z}$-modes, respectively. Equation (\ref{maineq}) allows us to get the electromagnetic energy $W_{em}$ radiated around $\omega_p$ in each mode.

Let us discuss results obtained by the model for a 2D
plasma radio source. Fig. \ref{fig3} shows the time variations
of $W_{em}$ in the modes $\mathcal{O}$, $\mathcal{X}$ and $\mathcal{Z}$, for $6$ sets of parameters, i.e.
with $\omega _{c}/\omega _{p}=0.05,$ three values of $\Delta N=0.02,$ $0.03,$
and $0.04,$ and two values of $c/v_{T}=c_{L}$. One observes that
in all cases the saturation of $W_{em}$ is not reached, due to the
non-self-consistency of the model. At late times, its variation can be
approximated with good accuracy by a linear growth, the slope of which
represents the time independent radiation rate $dW_{em}/dt$ at
frequency $\omega _{p}$. For $\omega _{c}/\omega _{p}=0.05\ $and $\Delta
N=0.02$ (upper row), the energy carried by the $\mathcal{X}$-mode is almost 
two orders of magnitude smaller than
 those of the $\mathcal{O}$- and $\mathcal{Z}$-modes;
in addition, the latter exhibits the highest radiation rate and asymptotic energy level.
As shown below using a third approach, $\mathcal{X}$-mode waves
can be radiated only if $\omega _{c}/\omega _{p}\lesssim \alpha \Delta N$,
where $\alpha \sim 1-2$ is a phenomenological parameter ($\alpha \sim 1$ for
above PIC simulations and $\alpha \sim 2$ for the present model). For $\Delta N=0.03$
(Fig. \ref{fig3}, middle row), the condition $\omega _{c}/\omega _{p}\lesssim \alpha \Delta N$
is fulfilled, and $\mathcal{X}$-mode waves are radiated. For $\Delta N=0.04$
(Fig. \ref{fig3}, bottom row), $\omega _{c}/\omega _{p}<\alpha \Delta N$\ is satisfied and $%
\mathcal{X}$-mode radiation is larger than for $\Delta N=0.03$, as expected.
These findings are in agreement with the results of PIC simulations presented
above (with close parameters), as well as with the analytical calculations
presented below.

\begin{figure}
    \centering
    \includegraphics[width=0.9\linewidth]{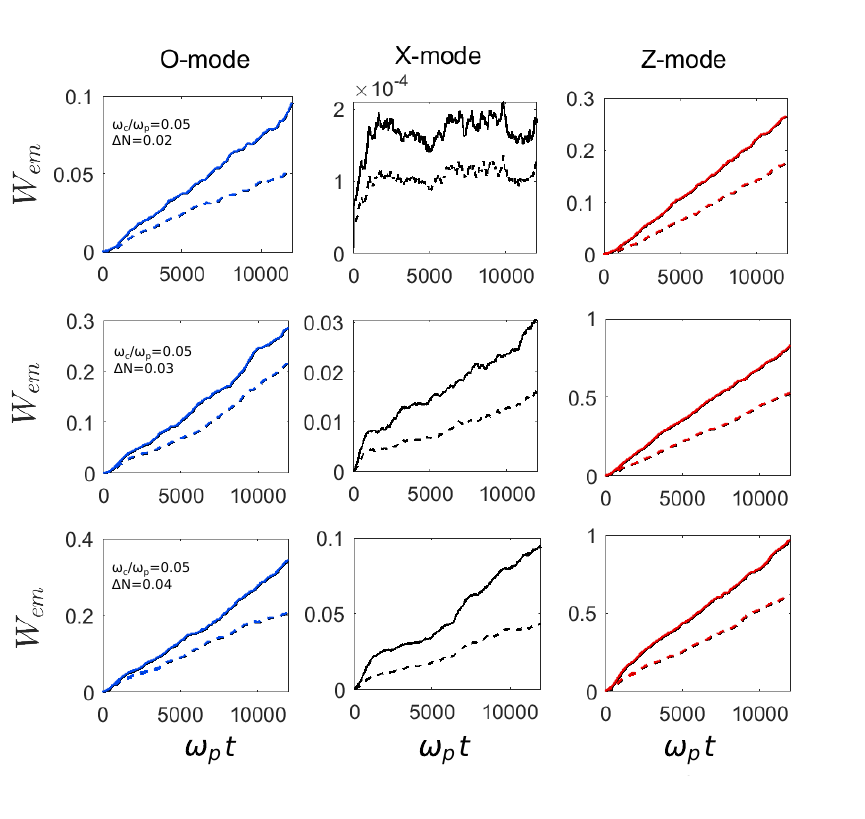}
    \caption{Variations with normalized time $\omega _{p}t$ of the
electromagnetic energy $W_{em}$ radiated around $\omega _{p}$ in the $\mathcal{O}$-, $\mathcal{X}$- and $%
\mathcal{Z}$-modes (left, middle and right columns, respectively), in linear
scales and arbitrary units, for $\omega _{c}/\omega _{p}=0.05$ and $\Delta N=0.02\ $(upper row), $%
0.03\ $(middle row), and $0.04\ $(bottom row), with two values of $c_{L}=40$
(solid lines) and $c_{L}=50$ (dashed lines).}
    \label{fig3}
\end{figure}

\begin{figure}
    \centering
    \includegraphics[width=0.9\linewidth]{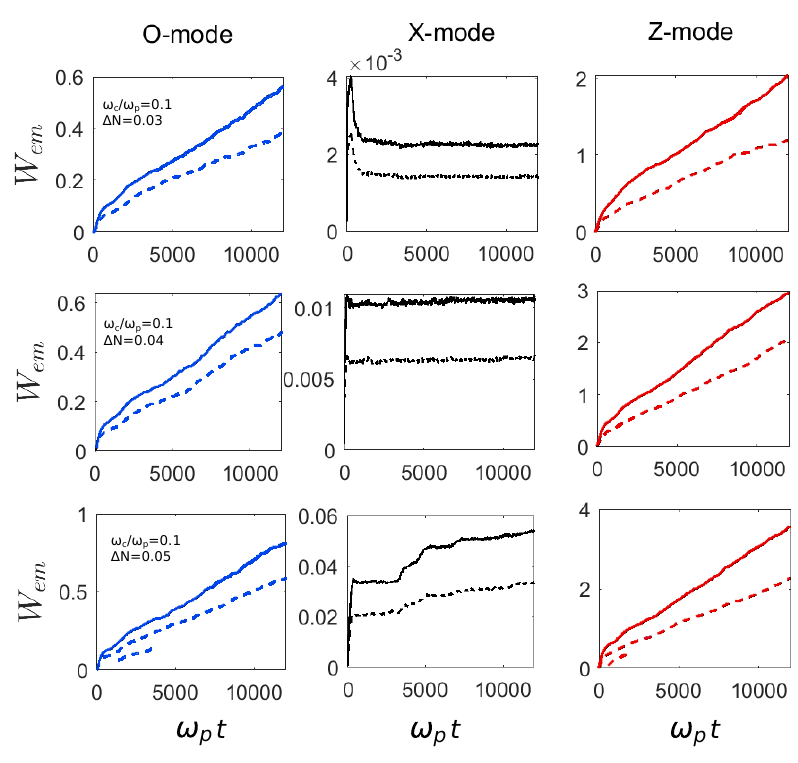}
    \caption{Variations with normalized time $\omega _{p}t$ of the
electromagnetic energy $W_{em}$ radiated around $\omega _{p}$ in the $\mathcal{O}$-, $\mathcal{X}$- and $%
\mathcal{Z}$-modes (left, middle and right columns, respectively), in linear
scales and arbitrary units, for $\omega _{c}/\omega _{p}=0.1$ and $\Delta N=0.03\ $(upper row), $%
0.04\ $(middle row), and $0.05\ $(bottom row), with two values of $c_{L}=40$
(solid lines) and $c_{L}=50$ (dashed lines).}
    \label{fig4}
\end{figure}
\begin{figure}
    \centering
    \includegraphics[width=0.9\linewidth]{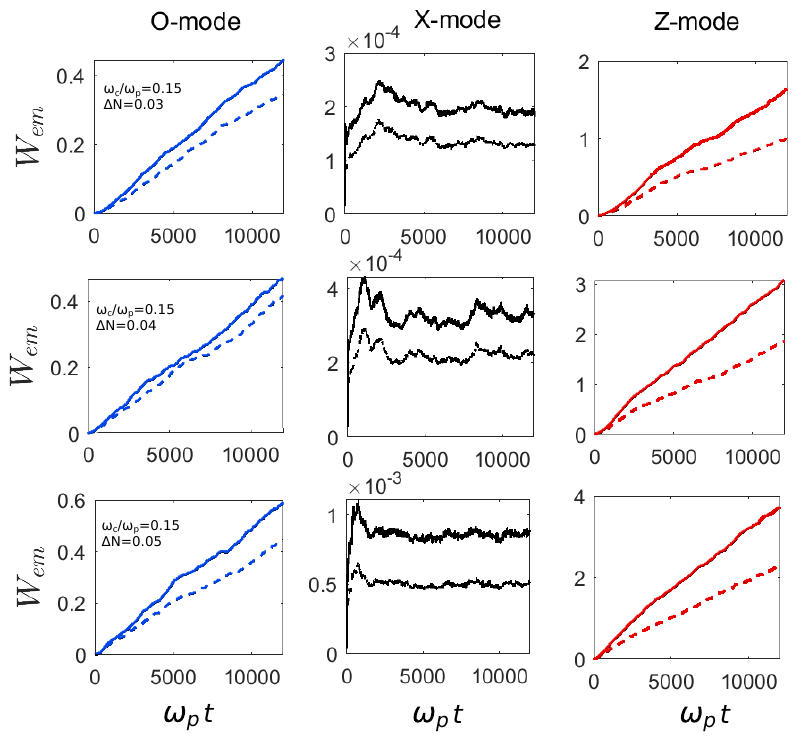}
    \caption{Fig. \ref{fig5}. Variations with normalized time $\omega _{p}t$ of the
electromagnetic energy $W_{em}$ radiated around $\omega _{p}$ in the $\mathcal{O}$-, $\mathcal{X}$- and $%
\mathcal{Z}$-modes (left, middle and right columns, respectively), in linear
scales and arbitrary units, for $\omega _{c}/\omega _{p}=0.15$ and $\Delta N=0.03\ $(upper row), $%
0.04\ $(middle row), and $0.05\ $(bottom row), with two values of $c_{L}=40$
(solid lines) and $c_{L}=50$ (dashed lines).}
    \label{fig5}
\end{figure}

Fig. \ref{fig4} shows the same type of plots as in  Fig. \ref{fig3} but for a larger plasma magnetization $\omega _{c}/\omega
_{p}=0.1,$ with $\Delta N=0.03,$ $0.04,$ and $0.05$ as well as two values of  $c_{L}$.
The same conclusions can be drawn as above. For $\Delta N=0.03\ $and $\Delta
N=0.04$ (upper and middle rows), $\mathcal{X}$-mode wave energy is roughly two orders of magnitude lower than those of the two other modes as $%
\omega _{c}/\omega _{p}>\alpha \Delta N$, whereas $\omega _{c}/\omega
_{p}\lesssim \alpha \Delta N$ ($\alpha \sim 2$) is satisfied for $\Delta
N=0.05,$ for which $\mathcal{X}$-mode waves are emitted. As observed above in Figs. \ref{fig2}-\ref{fig3}, $\mathcal{Z}$%
-mode waves exhibit the highest energies $W_{em}$ and radiation rates $dW_{em}/dt$ ($10^{-4}\lesssim dW_{em}/dt \lesssim 10^{-3}$). Finally, Fig. \ref{fig5} confirms all
previous statements for $\omega _{c}/\omega _{p}=0.15$ and $0.03\leq \Delta
N\leq 0.05$;  $\mathcal{X}$-mode waves are excited at negligible levels, with $\mathcal{Z}$-mode waves' energies and radiation rates exceeding
significantly those of the $\mathcal{O}$-mode, as in Figs. \ref{fig3} and %
\ref{fig4}.

The concordant results provided by the two different approaches presented above are
corroborated by analytic calculations performed in 3D geometry in the
framework of weak turbulence theory extended to the case of randomly
inhomogeneous plasmas, as demonstrated in detail in the section "Methods". It is shown that, starting from the differential equation (\ref{maineq}), a compact expression is obtained for the radiation rates $\mu^\pm$ of $\mathcal{X}$- and $\mathcal{Z}$-mode wave magnetic energies, in the form of an integral on $\mathcal{LZ}$ wavevectors 
\begin{equation}
\begin{split}
    &\mu^\pm=\frac{d}{\omega_{p}dt}
    {\displaystyle\int_V}
    \frac{d^{3}\mathbf{k}}{(2\pi)^{3}}\left\langle \left\vert B_{\mathbf{k}}^{\pm
    }(t)\right\vert ^{2}\right\rangle \simeq\frac{V}{\pi\lambda_{D}^{3}}\left(
    \frac{v_{T}}{c}\right)  ^{3}\times\\
    &\times    {\displaystyle\int_V}
    \frac{d^{3}\mathbf{k}_{2}}{(2\pi)^{3}}\left\vert \frac{\delta n_{-\mathbf{k}
    _{2}}\left(  t\right)  }{n_{0}}\right\vert ^{2}\left\vert E_{\mathbf{k}_{2}
    }(t)\right\vert ^{2}(3k_2^2\lambda_D^2\mp\frac{\omega_{c}
    }{\omega_{p}})^{3/2} \left(  \sin^{2}\theta_{2}\left(  1\pm
    2\frac{\omega_{c}}{\omega_{p}}\right)  +\frac{\cos^{2}\theta_{2}}{15}\right)
\end{split}
    \label{ratefinal22}
    \end{equation}
with     
\begin{equation}
k_{\pm }^{2}\lambda_{D}^{2}=\frac{2v_T^2}{c^2}\left(  3k_{2}^{2}\lambda_{D}^{2}\mp\frac{\omega_{c}
    }{\omega_{p}}\right)>0,
\end{equation}
and $\cos{\theta_2}=\mathbf{k_2} \cdot \mathbf{h}/k_2$; $\mathbf{k}_2$ is the wavevector of $\mathcal{LZ}$ waves; the rhs integral in equation (\ref{ratefinal22}) includes the spectral energy  $|E_{\mathbf{k_2}}(t)|^2$ of the turbulent $\mathcal{LZ}$ waves' spectrum  and the plasma characteristics, i.e. the applied density fluctuations' spectrum $|\delta n_{-\mathbf{k_2}}/n_0|^2$ of average level $\Delta N$, the electron thermal velocity $v_T/c$, and the magnetization $\omega_c/\omega_p$. Note that the radiation rates scale as $(v_T/c)^3$ and increase when approaching the Sun.

Then, knowing the spectra of turbulent $\mathcal{LZ}$ waves and density fluctuations at large times, equation (\ref{ratefinal22}) enables to numerically calculate the radiation rates of electromagnetic $\mathcal{X}$- and $\mathcal{Z}$-mode waves. The condition $k_{\pm }^{2}>0$ is required, what is always fulfilled for the $\mathcal{Z}$-mode ($k_{-}^{2}>0$ as $3k_{2}^{2}%
\lambda_{D}^{2}+\omega _{c}/\omega _{p}>0$) but not for the $\mathcal{X}$%
-mode ($k_{+}^{2}$ is negative if  $3k_{2}^{2}\lambda _{D}^{2}<\omega _{c}/\omega _{p}$), explaining why the latter can not be radiated under some plasma conditions depending
on $\omega _{c}/\omega _{p}$ and  $\Delta N$. Indeed, for electromagnetic radiation at $\omega
_{p}$ via linear mode conversion at constant frequency, we can write that $%
3k_{2}^{2}\lambda _{D}^{2}\sim \alpha \Delta N,$ where $\alpha \sim 1-2$
depending  on the two approaches used above. For $\omega _{c}/\omega _{p}\gtrsim
\alpha \Delta N$, $\mathcal{Z}$-mode waves are radiated but not $\mathcal{X}
$-mode ones,
in agreement with the PIC simulations' and the model's results shown in Figs. %
\ref{fig1}-\ref{fig2} and Figs. \ref{fig3}-\ref{fig5}, respectively. 
Then, contrary to the other modes, $\mathcal{X}$-mode waves can only be
emitted at significant energy levels by radio sources with $\omega _{c}/\omega _{p}\lesssim \alpha
\Delta N$.

In addition,
equation (\ref{ratefinal22}) shows that $\mathcal{X}$- and $\mathcal{Z}$%
-mode waves can reach similar radiation rates in the same source when $\omega
_{c}/\omega _{p}\ll \Delta N$ but that, for $\omega _{c}/\omega _{p}<\Delta
N,$ the energy carried by $\mathcal{X}$-mode waves should be noticeably to significantly less
than that of $\mathcal{Z}$-mode waves, as verified by the two above
approaches. At solar wind conditions near 1 AU where $\Delta N\lesssim 0.02$
can be reasonably expected, $\mathcal{X}$-mode waves can be radiated only by very weakly magnetized
radio sources. In coronal plasmas, where $\Delta
N$ and $\omega_c/\omega_p$ can reach larger values up to  $0.1$ and $0.5$ (and even more), respectively, $\mathcal{X}$-mode waves can not be radiated at $%
\omega _{p}$ by mode conversion of electrostatic wave turbulence generated 
by electron beams in inhomogeneous plasma sources during Type III radio bursts, 
except if they can result from other types of conversion processes of $\mathcal{O}$-mode or 
even $\mathcal{Z}$-mode waves or from other unknown mechanisms.

\section{Discussion}
For the first time, electromagnetic radiation at $\omega_p$ by electrostatic upper-hybrid ($\mathcal{LZ}$) wave
turbulence generated by electron beams in randomly inhomogeneous and weakly
to moderately magnetized plasmas has been determined. The independent
approaches presented above have demonstrated that electromagnetic emissions at $\omega_p$ are mainly 
radiated in the $\mathcal{Z}$-mode, and less in the $\mathcal{O}$- and $%
\mathcal{X}$-modes, which are commonly observed by satellites and
radiotelescopes during Type III solar radio bursts.
In addition, PIC simulations show that this statement remains true at least up
to $\omega _{c}/\omega _{p}=0.5$, i.e. down to distances around $10$
solar radii ($\sim0.05$ AU) to the Sun. Except for very weakly magnetized plasmas with $\omega _{c}/\omega _{p}\lesssim0.02$, $\mathcal{O}$-mode 
energy is generally an order of magnitude lower than $\mathcal{Z}$%
-mode wave energy, and significantly higher than $\mathcal{X}$-mode energy.
Our results differ from those obtained by previous works on linear mode
conversion of monochromatic Langmuir or upper-hybrid waves interacting with a fixed density gradient (and
not varying random density fluctuations), which found that only $\mathcal{O}$- and $%
\mathcal{X}$-mode emissions are radiated (e.g. \cite{Yin1998,Kim2008,Kim2013,Schleyer2013,Schleyer2014}). Interactions between weak electrostatic high-frequency wave turbulence generated by beam-plasma instability and low-frequency density turbulence inherent to solar wind plasmas, as modeled in this work, are indeed the cause of electromagnetic radiation at $\omega_p$.

Another important point revealed by the three independent approaches 
presented above concerns the possible lack
of $\mathcal{X}$-mode radiation, depending on the ratio between the
magnetization $\omega _{c}/\omega _{p}$ and the average level of
density fluctuations $\Delta N$. It is shown that, when $\omega _{c}/\omega
_{p}\gtrsim \alpha \Delta N$ ($\alpha \sim 1-2$),  $\mathcal{X}$-mode waves
cannot be radiated; when $\omega _{c}/\omega _{p}\sim \alpha \Delta N$, the
radiation rate in this mode can become weakly positive and, when $\omega _{c}/\omega
_{p}\lesssim \alpha \Delta N,$ these waves can be emitted with significant
amplitudes, that can even reach those of $\mathcal{Z}$-mode waves. The presence or lack of $\mathcal{X}$-mode radiation can also be used to locate Type III radio sources.
The condition $\omega _{c}/\omega _{p}\gtrsim \alpha
\Delta N$ being easily satisfied in solar wind regions with $\omega _{c}/\omega _{p}\gtrsim 0.1-0.2$, Type III radio sources at distances to the Sun below 
roughly 10 solar radii should rarely emit $\mathcal{X}$%
-mode radiation by mode conversion of $\mathcal{LZ}$ wave turbulence. However, other processes producing $\mathcal{X}$-mode radiation can not be a priori excluded. On the other
hand, in the solar wind near 1 AU, $\mathcal{X}$-mode waves can be radiated with significant energies only by
very weakly magnetized radio sources with $\omega_{c}/\omega_{p}\lesssim
0.02$, which can be found near 1 AU.

Our findings show that most of electromagnetic energy emitted by
electrostatic wave turbulence in Type III radio sources, which consists in $%
\mathcal{Z}$-mode waves with frequencies $\omega \lesssim \omega _{p}$, can
not escape far from its region of generation, due to dispersive and propagation
characteristics (as group velocity), and thus cannot be
observed far away from it. They should nowadays be detectable  near Type III radio sources by satellites
as Parker Solar Probe and Solar Orbiter which approach the Sun at distances never reached before. As, according to our studies, only a
small part of electromagnetic energy radiated at $\omega_p$ (roughly $10\%$) can be
actually observed in the form of $\mathcal{O}$-mode waves, and more rarely,
of $\mathcal{X}$-mode waves, 
only partial information on the mechanisms of generation of
electromagnetic emissions within radio sources is accessible at far distances from them. This is a further justification of the importance of this work.

In addition, the question arises of how energy radiated in the electromagnetic $\mathcal{Z}$-mode, 
which remains inside or close to the radio source, is
further transformed and dissipated. In randomly inhomogeneous plasmas,
$\mathcal{LZ}$ waves tend to be localized over time within density
depletions, where they can become $\mathcal{Z}$-mode waves, which can be amplified when their group velocity decreases
to vanishing values near their cutoff frequency. If their amplitudes significantly exceed
those of $\mathcal{O}$-mode waves and their frequencies are close to $\omega_p$, 
they can coalesce into harmonic waves radiated near frequency $2\omega _{p}$, that can freely escape outside the
source. Moreover, as the radio source moves with the beam, $\mathcal{Z}$-mode waves can encounter various density gradients and, in such circumstances, can be
converted into $\mathcal{O}$-mode waves. Such processes are however beyond the scope of this work.

Our results show that most of electromagnetic energy radiated at $\omega _{p}
$ in radio sources and capable of escaping from them to infinity is more generally
carried by $\mathcal{O}$-mode waves than by $\mathcal{X}$-mode ones (except in very weakly magnetized plasmas near Earth). As a
consequence, observations performed by satellites in the solar wind should
mostly reveal $\mathcal{O}$-mode type polarization, at least in the vicinity
of sources. Far away from them, the impact of density inhomogeneities as
gradients or random fluctuations of various scales on the propagation of  $%
\mathcal{O}$-mode emissions can depolarize them. In this view, actual
observations performed during Type III radio bursts report mostly $\mathcal{O%
}$-mode type polarization affected by depolarization effects (e.g. \cite{Jebaraj2023a,Pulupa2020}), usually attributed to the
simultaneous emission of $\mathcal{X}$-mode waves. Our results
suggest that, when plasma conditions do not allow $\mathcal{X}$-mode
waves to be generated, $100\%$ of $\mathcal{O}$-mode polarization should be
observed, as reported recently for some events \cite{Jebaraj2023a}. On the contrary, when $\omega _{c}/\omega _{p}\lesssim \alpha \Delta N$, $\mathcal{X}$-mode waves radiated together with $\mathcal{O}$-mode ones can
depolarize them. Such effects could also be due to the possible conversion
of a part of $\mathcal{O}$-mode waves into $\mathcal{X}$-mode ones during their
propagation away from the source along large-scale density
gradients encountered during their travel from Sun to Earth.

\section{Methods}
\subsection{2D/3V PIC simulations}
Simulations have been performed with the open source relativistic
full particle code SMILEI \cite{Derouillat2018}, which includes both electrons and ions as finite-size particles (Particle-In-Cell code). It follows the Vlasov-Maxwell description
for a collisionless plasma and solves Maxwell and Poisson equations using a Yee mesh with centered electric and magnetic fields according to the
finite-difference time-domain (FDTD) method. The integration of particle motion is performed using the Boris scheme. SMILEI has an hybrid MPI-OpenMP programmation scheme and was ported on GPU architecture. In particular, the 
computation of the current has been accelerated with a CUDA kernel while other subroutines use either OpenAcc or OpenMP Pragmas with high-level compiler directives.

We use in this work the GPU 2D/3V version
of SMILEI, corresponding to two-dimensional Cartesian coordinates in space, while macroparticles have three velocity components. To achieve a sufficiently realistic description of the phenomena under study and include self-consistently in the required frequency and wavelength domains, numerical constraints have to be satisfied. Indeed, 
the processes described in this paper include both electrostatic and electromagnetic waves, whose wavelengths span more than two orders of magnitude, requiring large simulation boxes, not to mention the fact that only high spatial resolution allows to discriminate between the different electromagnetic modes when plasma magnetization is low, as in the solar wind. Finally, waves involved in electromagnetic emission mechanisms are both of low- and high-frequency, so both fast electron dynamics and slow ion dynamics need to be tracked, requiring very long simulation times. 

Exceptionnally high wavenumber and frequency resolutions up to $\delta k\simeq 0.001\lambda _{D}^{-1}$ and $\delta \omega =0.002\omega _{p}$
 are used to identify the  electromagnetic modes $\mathcal{O%
}$, $\mathcal{X}$ and $\mathcal{Z}$ and to separate them one from the other near the frequency $\omega_p$.
Simulation boxes with sizes up to $L_{x}\times L_{y}=5792\times 2896\lambda
_{D}^{2}\simeq116\times58c^2/\omega_p^2$ are used ($\lambda _{D}$ is the electron Debye length). Physical conditions are relevant to
Type III radio bursts in the solar wind, with $0.02\leq \omega
_{c}/\omega _{p}\leq 0.5$ (corresponding to distances to Sun down to $\sim0.05$ AU), $v_{b}\simeq 13v_{T}\simeq 0.25c$ and $n_{b}/n_{0}=5 \cdot 10^{-4}$ (beam velocity and relative density, where $n_{0}$ is the
average plasma density), as well as $T_{e}/T_{i}=10$ and $m_{e}/m_{i}=1/1836$
(electron-to-ion temperature and mass ratios). 

The plasma is initially set 
inhomogeneous, with applied random density fluctuations $\delta n$ of average levels $\Delta
N={<(\delta n/n_0)^2>}^{1/2}\lesssim 0.05$ and wavelengths much larger than those of $\mathcal{LZ}$ waves \cite{Celnikier1987}. Note that 1800 particles per cell and per species (beam electrons, plasma electrons, and plasma ions) are used, in order to reduce the numerical noise to very low levels below $\Delta N,$ i.e. to less than 2.5$\%$.  More details on numerical parameters can be found in our previous works \cite{KrafftSavoini2022a,KrafftSavoini2021,KrafftSavoini2024,Krafft2024,KrafftSavoini2023}.

\subsection{Theoretical model}
The dynamics of $\mathcal{LZ}$ wave turbulence in randomly inhomogenous and weakly magnetized ($\omega _{c}/\omega _{p}\lesssim 0.2$) radio sources is calculated using modified Zakharov equations involving
weak magnetic effects  \cite{KrasnoselskikhSotnikov1977}, where we neglect ponderomotive terms, especially since strong turbulence phenomena are generally not occurring in the solar wind and because, in plasmas with random density fluctuations, electromagnetic radiation by upper-hybrid wave turbulence transformations on inhomogeneities (as scattering or conversion) is dominant with respect to three-wave nonlinear interactions \cite{KrafftSavoini2022a}. Then we can suppose that low-frequency waves evolve linearly, and the corresponding non-self-consistent system of equations can be written as follows 
\begin{equation}
\nabla ^{2}\left( i\frac{\partial \tilde{\varphi}}{\partial t}+\frac{3\omega
_{p}}{2}\lambda _{D}^{2}\nabla ^{2}\tilde{\varphi}\right) -\frac{\omega
_{c}^{2}}{2\omega _{p}}\nabla _{\perp }^{2}\tilde{\varphi}\simeq \frac{%
\omega _{p}^{2}}{2\omega }\nabla \cdot \left( \frac{\delta n}{n_{0}}\nabla 
\tilde{\varphi}\right) ,\text{ \ \ \ \ \ \ }\left( \frac{\partial
^{2}}{\partial t^{2}}-c_{s}^{2}\Delta \right) \frac{\delta n}{n_{0}}\simeq 0,
\label{hf-eq}
\end{equation}%
where $c_{s}=\left( \left( T_{e}+3T_{i}\right) /m_{i}\right) ^{1/2}$ is the
ion acoustic velocity; $\tilde{%
\varphi}\mathbf{\mathbf{(r,}}t\mathbf{\mathbf{)}}$ is the slowly varying
envelope of the upper-hybrid waves' potential; $\nabla _{\perp }$ is the gradient perpendicular to $\mathbf{B}_0$. The current $\delta \mathbf{%
\mathbf{j(r,}}t\mathbf{\mathbf{)}}$ generated at each time $t$ and position $%
\mathbf{\mathbf{r}}$ in the plasma source is calculated by numerically
solving equations (\ref{hf-eq}). It generates electromagnetic emissions at frequency $\omega _{p}$, which are supposed to leave their turbulent source and to propagate away to infinity through a uniform plasma. 

Combining Maxwell equations, we get that%
\begin{equation}
\left( \frac{\partial ^{2}}{\partial t^{2}}-c^{2}\nabla ^{2}\right) \mathbf{B%
}=4\pi c\mathbf{\nabla }\times \left( \delta \mathbf{J}+\delta \mathbf{%
\mathbf{j}}\right) \mathbf{,}  \label{Maxwell}
\end{equation}%
where $\mathbf{B}(\mathbf{r},t)$ is the wave magnetic field; the current density $\delta 
\mathbf{J}=-en_{0}\mathbf{v}_{e}$, which controls wave dispersion, is generated by electrons (of charge $%
-e<0$) moving with velocity $\mathbf{v}_{e}$ in the magnetized plasma of
density $n_{0}$. The external current density $\delta \mathbf{j}=-e\delta n%
\mathbf{v}_{e}\mathbf{,}$ due to scattering of $\mathcal{LZ}$ waves on density
fluctuations $\delta n$, oscillates at a frequency close to $\omega _{p}$ as 
$\omega _{c}\ll \omega _{p}$. We neglect here the non-potential part of
$\mathcal{LZ}$ electric fields, whose envelope can then be written as $
\mathbf{E}\simeq -\nabla \tilde{\varphi}$. In the linear approximation, keeping only
lowest order terms in $\omega _{c},$ we get the slowly varying envelope of
the external current density
\begin{equation}
4\pi \delta \mathbf{\mathbf{\tilde{j}}}\simeq \mathbf{-}\frac{i\omega
_{p}^{2}\omega }{\omega ^{2}-\omega _{c}^{2}}\frac{\delta n}{n_{0}}\left( 
\mathbf{\nabla }\tilde{\varphi}+i\frac{\omega _{c}}{\omega }\mathbf{h}\times 
\mathbf{\nabla }\tilde{\varphi}\right) ,  \label{Current}
\end{equation}%
where $\omega \simeq \omega _{p}$ is the radiation frequency; $\mathbf{h}$ is the unitary vector along $\mathbf{B}%
_{0}$. As different electromagnetic wave modes and polarizations exist around $\omega _{p}$, we describe them by introducing the vector 
$\mathbf{\alpha}_\mathbf{k}^{\pm}=\mathbf{k}_{\perp }\pm i\mathbf{h}\times \mathbf{k}_{\perp}$  directed accross $\mathbf{B_0}$, 
where $\mathbf{k}_{\perp }$ is the perpendicular component of the wavevector 
$\mathbf{k}$; the signs $+$ and $-$ correspond to the $\mathcal{X}$- and 
$\mathcal{Z}$-modes, respectively, which can be both excited around $\omega \simeq
\omega _{p}$ in weakly magnetized plasmas. The Fourier components $\mathbf{B}_{%
\mathbf{k}}^{\pm }(t)$ of the wave magnetic field envelope $\mathbf{\tilde{B}(r,}t)$ can be expressed as a function of their amplitudes $B_{\mathbf{k}}^{\pm }(t)$ as 
\begin{equation}
\mathbf{B}_{\mathbf{k}}^{\pm }=i%
\frac{\mathbf{k\times \mathbf{\alpha} }_{\mathbf{k}}^{\pm }}{\left\vert \mathbf{%
k\times \mathbf{\alpha} }_{\mathbf{k}}^{\pm }\right\vert }B_{\mathbf{k}}^{\pm }=i%
\mathbf{a}^{\pm }_{\mathbf{k}}B_{\mathbf{k}}^{\pm }.  \label{ak}
\end{equation}%
Then, introducing $\delta \mathbf{\tilde{j}}^{\pm }=\delta 
\mathbf{\tilde{j}}\pm i\mathbf{h}\times \delta \mathbf{\tilde{j}}$ and
separating slowly varying envelopes from fast oscillating phases at $%
\omega _{p},$ equation (\ref{Maxwell}) can be written in $\mathbf{k}$-space
as%
\begin{equation}
\left( i\frac{\partial }{\partial t}-\Delta \omega _{\pm }\right) B_{\mathbf{%
k}}^{\pm }\simeq-\frac{2\pi c}{\omega _{p}}\mathbf{a}^{\pm\ast }_{\mathbf{k}}\cdot 
\mathbf{k}\times \left( \delta \mathbf{j}_{\mathbf{k}}\pm i\mathbf{h}\times
\delta \mathbf{j}_{\mathbf{k}}\right) ,  \label{maineq2}
\end{equation}%
where the superscript '$\ast $' indicates complex conjugation; $\delta 
\mathbf{j}_{\mathbf{k}}(t)$ is the Fourier component of $\delta \mathbf{\tilde{j%
}}(\mathbf{r},t)$; $\Delta \omega _{\pm }$ is the frequency detuning and $\omega _{\mathbf{%
k}}^{\pm }$ is the dispersion relation of the $\mathcal{X}$- and $\mathcal{Z}
$-mode waves near their cutoff frequencies 
\begin{equation}
\Delta \omega _{\pm }=\omega _{\mathbf{k}}^{\pm }-\omega _{p},\ \ \ \ \ \omega _{\mathbf{k}}^{\pm }\simeq \omega _{p}\pm \frac{%
\omega _{c}}{2}+\frac{c^{2}\left( k^{2}+k_{\parallel }^{2}\right) }{4\omega
_{p}},  \label{delta}
\end{equation}%
where $\mathbf{k}_{\parallel }$ is the parallel component of $\mathbf{k}$ with respect to $\mathbf{B}_0$.
Then, using the current $\delta \mathbf{j(r,}t\mathbf{)}$ calculated at each time and position using equation (\ref%
{hf-eq}),
one can compute $\delta \mathbf{j}_{%
\mathbf{k}}$ as well as the rhs term of (\ref{maineq2}), and further
numerically solve this equation to obtain the magnetic energy $\left\vert B_{\mathbf{k%
}}^{\pm }\right\vert ^{2}$ and its time derivative, i.e. the radiation rate
$\mu^{\pm}$ at frequency $\omega _{p}$. Note that equation (\ref{maineq2}) is also valid
for the $\mathcal{O}$-mode, with however other expressions for its
rhs term, the frequency detuning $\Delta \omega _{\pm },$ the
dispersion $\omega _{\mathbf{k}}^{\pm}$ and the polarization $\mathbf{a}^{\pm }_{%
\mathbf{k}},$ that will be presented in detail in a forthcoming paper; we however
present above its energy and radiation rate together with those of the $%
\mathcal{X}$- and $\mathcal{Z}$-modes.

\subsection{Analytical determination of electromagnetic radiation rates}

Starting from equation (\ref{maineq}), 
we get below the radiation rates of electromagnetic modes' magnetic energy,
in the form of a compact and tractable expression (\ref{ratefinal2}) (also labelled with   (\ref{ratefinal22}) in the main text).
Combining equations (\ref{maineq}) and  (\ref{Current}),
we can write that 
\begin{equation}
\left( i\frac{\partial }{\partial t}-\Delta \omega _{\pm }\right) B_{\mathbf{%
k}}^{\pm }\simeq-\frac{c}{2}\frac{\omega _{p}^{2}}{\omega _{p}^{2}-\omega _{c}^{2}%
}\sum_{\mathbf{k=k}_{1}+\mathbf{k}_{2}}\mathbf{a}_{\mathbf{k}}^{\pm\ast }\cdot
\eta _{\mathbf{kk}_{2}}\frac{\delta n_{\mathbf{k}_{1}}}{n_{0}}\varphi _{%
\mathbf{k}_{2}},  \label{akG}
\end{equation}%
where $\delta n_{\mathbf{k}_{1}}$and $\varphi _{\mathbf{k}_{2}}$ are the
Fourier components of $\delta n$ and $\tilde{\varphi}$, respectively, and%
\begin{equation}
\eta _{\mathbf{kk}_{2}}=\mathbf{k}\times \left[ \left( 1\pm i\mathbf{h}%
\times \right) \left( \mathbf{k}_{2}+i\frac{\omega _{c}}{\omega }\mathbf{h}%
\times \mathbf{k}_{2}\right) \right] ,  \label{etakk2}
\end{equation}%
where the wavevectors $\mathbf{k}_{1}$, $\mathbf{k}_{2}$ and $\mathbf{k}$ correspond to
low- and high-frequency (ion acoustic and $\mathcal{LZ}$) electrostatic
waves as well as to electromagnetic waves, respectively. At the large times considered here (as explianed in the section "Results"), we can neglect the
small initial values $B^{\pm}_{\mathbf{k}}(0)$ so that, by integrating (\ref{akG}%
) and squarring, we get the spectral magnetic energy 
\begin{equation}
\begin{split}
&\left\vert B_{\mathbf{k}}^{\pm }(t)\right\vert ^{2}\simeq \frac{c^{2}}{4}%
\left( \frac{\omega _{p}^{2}}{\omega _{p}^{2}-\omega _{c}^{2}}\right)
^{2}\sum_{\mathbf{k=k}_{1}+\mathbf{k}_{2}}\sum_{\mathbf{k=k}_{3}+\mathbf{k}%
_{4}}\Bigg[(\mathbf{a}_{\mathbf{k}}^{\pm\ast }\cdot \eta _{\mathbf{kk}_{2}})(\mathbf{a}_{\mathbf{k}}^\pm\cdot \eta _{\mathbf{kk}_{4}}^{\ast })\times\\
&\times \int_{0}^{t}\int_{0}^{t}\frac{\delta n_{\mathbf{k}_{1}}\left(
t^{\prime }\right) }{n_{0}}\frac{\delta n_{\mathbf{k}_{3}}^{\ast }\left(
t^{\prime \prime }\right) }{n_{0}}\varphi _{\mathbf{k}_{2}}(t^{\prime
})\varphi _{\mathbf{k}_{4}}^{\ast }(t^{\prime \prime })e^{i\Delta \omega
_{\pm }\left( t^{\prime }-t^{\prime \prime }\right) }dt^{\prime }dt^{\prime
\prime }\Bigg].
\end{split}
\label{Bk22}
\end{equation}
The energy $\left\vert B_{\mathbf{k}}^{\pm }(t)\right\vert ^{2}$ experiences
statistical fluctuations and its growth with time on average provides its
radiation rate $\mu^{\pm}$ (see also the section "Results"). As the dependence with time of random density fluctuations'
amplitudes is weak, one can neglect the small phases of low-frequency
oscillations and assume that $\left\langle \delta n_{\mathbf{k}_{1}}\left(
t^{\prime }\right) \delta n_{\mathbf{k}_{3}}^{\ast }\left( t^{\prime \prime
}\right) \right\rangle \simeq \delta _{\mathbf{k}_{1},\mathbf{k}_{3}}\left\vert
\delta n_{\mathbf{k}_{1}}\right\vert ^{2}$, where brackets denote
statistical averaging, supposed to be consistent with time averaging. Then  
\begin{equation}
\begin{split}
&\left\langle \left\vert B_{\mathbf{k}}^{\pm }(t)\right\vert
^{2}\right\rangle \simeq \frac{c^{2}}{4}\left( \frac{\omega _{p}^{2}}{\omega
_{p}^{2}-\omega _{c}^{2}}\right) ^{2}\sum_{\mathbf{k=k}_{1}+\mathbf{k}%
_{2}}\Bigg[\left\vert \mathbf{a}_{\mathbf{k}}^{\pm\ast }\cdot \eta _{\mathbf{kk}%
_{2}}\right\vert ^{2}\left\vert \frac{\delta n_{\mathbf{k}_{1}}\left(
t\right) }{n_{0}}\right\vert ^{2}\times\\
&\times\left\langle
\int_{0}^{t}\int_{0}^{t}\varphi _{\mathbf{k}_{2}}(t^{\prime })\varphi _{%
\mathbf{k}_{2}}^{\ast }(t^{\prime \prime })e^{i\Delta \omega _{\pm }\left(
t^{\prime }-t^{\prime \prime }\right) }dt^{\prime \prime }dt^{\prime
}\right\rangle\Bigg] ,  
\end{split}
\label{avenergy}
\end{equation}
where the slowly varying term $\left\vert \delta n_{\mathbf{k}_{1}}\left(
t\right) \right\vert ^{2}$ has been taken out of time integrals. To go further
we suppose, as in weak turbulence theory, that correlations between
amplitudes of upper-hybrid waves decay exponentially as $\left\langle
\varphi _{\mathbf{k}_{2}}(t^{\prime })\varphi _{\mathbf{k}_{2}}^{\ast
}(t^{\prime \prime })\right\rangle =\left\vert \varphi _{\mathbf{k}%
_{2}}\right\vert ^{2}\exp \left( -\nu _{\mathbf{k}_{2}}\left\vert t^{\prime
}-t^{\prime \prime }\right\vert -i\delta \omega _{\mathbf{k}_{2}}\left(
t^{\prime }-t^{\prime \prime }\right) \right) $, where $\delta \omega _{%
\mathbf{k}_{2}}=\omega _{\mathbf{k}_{2}}-\omega _{p}$. The frequency $\nu _{\mathbf{k}%
_{2}}>0$ is determined by the interactions between $\varphi _{\mathbf{k}_{2}}
$ and $\delta n_{\mathbf{k}_{1}}$. At large times $t,$ and if $\nu _{\mathbf{%
k}_{2}}$ is not too large, the double integral in (\ref{avenergy}) tends to $2\pi t\delta \left( \Delta \omega _{\pm }-\delta \omega _{%
\mathbf{k}_{2}}\right) $, where $\delta$ is the Dirac function, so that we get
\begin{equation}
\frac{d}{dt}\left\langle \left\vert B_{\mathbf{k}}^{\pm }(t)\right\vert
^{2}\right\rangle \simeq \frac{\pi c^{2}}{2}\left( \frac{\omega _{p}^{2}}{%
\omega _{p}^{2}-\omega _{c}^{2}}\right) ^{2}\sum_{\mathbf{k=k}_{1}+\mathbf{k}%
_{2}}\left\vert \mathbf{a}_{\mathbf{k}}^{\pm\ast }\cdot \eta _{\mathbf{kk}%
_{2}}\right\vert ^{2}\left\vert \frac{\delta n_{\mathbf{k}_{1}}\left(
t\right) }{n_{0}}\right\vert ^{2}\left\vert \varphi _{\mathbf{k}%
_{2}}(t)\right\vert ^{2}\delta \left( \omega _{\mathbf{k}}^{\pm }-\omega _{%
\mathbf{k}_{2}}\right) ,  \label{rate1}
\end{equation}%
where we used that $\Delta \omega _{\pm }-\delta \omega _{\mathbf{k}_{2}}=\omega _{%
\mathbf{k}}^{\pm }-\omega _{\mathbf{k}_{2}}$.
Then, taking into account that $\left\vert \mathbf{k}\right\vert \ll
\left\vert \mathbf{k}_{1}\right\vert ,\left\vert \mathbf{k}_{2}\right\vert $
and thus $\mathbf{k}_{2}\simeq -\mathbf{k}_{1},$ and assuming that waves'
and density fluctuations' spectra are sufficiently smooth, we can perform
the sums over $\mathbf{k}_{2}$ and $\mathbf{k}$ independently and obtain 
\begin{equation}
\begin{split}
&\frac{d}{dt}\sum_{\mathbf{k}}\left\langle \left\vert B_{\mathbf{k}}^{\pm
}(t)\right\vert ^{2}\right\rangle \simeq \frac{\pi c^{2}}{2}\left( \frac{%
\omega _{p}^{2}}{\omega _{c}^{2}-\omega _{p}^{2}}\right) ^{2}\times\\
&\times\sum_{\mathbf{k}%
_{2}}\left[\left\vert \frac{\delta n_{-\mathbf{k}_{2}}\left( t\right) }{n_{0}}%
\right\vert ^{2}\left\vert E_{\mathbf{k}_{2}}(t)\right\vert ^{2}\left( \sum_{%
\mathbf{k}}\left\vert \mathbf{a}_{\mathbf{k}}^{\pm\ast }\cdot \eta _{\mathbf{kk}%
_{2}}^{\prime }\right\vert ^{2}k^{2}\delta \left( \omega _{\mathbf{k}}^{\pm
}-\omega _{\mathbf{k}_{2}}\right) \right) \right],
\end{split}
\label{Bk1}
\end{equation}
where we defined $\eta _{\mathbf{kk}_{2}}^{\prime }$ as $\left\vert \mathbf{a%
}_{\mathbf{k}}^{\pm\ast }\cdot \eta _{\mathbf{kk}_{2}}\right\vert
^{2}=k^{2}k_{2}^{2}\left\vert \mathbf{a}_{\mathbf{k}}^{\pm\ast }\cdot \eta _{%
\mathbf{kk}_{2}}^{\prime }\right\vert ^{2}$, expressed $\left\vert
\varphi _{\mathbf{k}_{2}}(t)\right\vert ^{2}$ as a function of $\left\vert E_{\mathbf{k}_{2}}(t)\right\vert ^{2}$ and used that $|\delta n_{\mathbf{k}_1}|^2\simeq|\delta n_{-\mathbf{k}_2}|^2$.

Further, transforming discrete sums into continuous ones via the equivalence  $V_{s}^{-1}\sum_{\mathbf{k}}f(\mathbf{k})=\int_{V_s} f(\mathbf{%
k})d^{s}\mathbf{k/}(2\pi )^{s}$, where $V_{s}$ is a volume of
dimension $s$, $d^{s-1}\Omega $ is the differential on angles and $d^{s}\mathbf{k}=k^{s-2}dk^{2}d^{s-1}\Omega /2,$ we get the radiation rates $\mu^{\pm}$ in the form%
\begin{equation}
\begin{split}
        &\mu^\pm=\frac{d}{\omega_p dt}
        {\displaystyle\int_{V_s}}
        \frac{d^{s}\mathbf{k}}{(2\pi)^{s}}\left\langle \left\vert B_{\mathbf{k}}^{\pm
        }(t)\right\vert ^{2}\right\rangle \simeq\frac{V_{s}}{(2\pi)^{s}}\frac{\pi
        c^{2}}{4\omega_p}\left(  \frac{\omega_{p}^{2}}{\omega_{c}^{2}-\omega_{p}^{2}}\right)
        ^{2}\times \\
        &\times{\displaystyle\int_{V_s}}
        \frac{d^{s}\mathbf{k}_{2}}{(2\pi)^{s}}\left\vert \frac{\delta n_{-\mathbf{k}%
        _{2}}\left(  t\right)  }{n_{0}}\right\vert ^{2}\left\vert E_{\mathbf{k}_{2}%
        }(t)\right\vert ^{2}\int_{\Omega}\int_{0}^{\infty}\left\vert \mathbf{a}_{\mathbf{k}}%
        ^{\pm\ast}\cdot\eta_{\mathbf{kk}_{2}}^{\prime}\right\vert ^{2}\delta\left(
        \omega_{\mathbf{k}}^{\pm}-\omega_{\mathbf{k}_{2}}\right)  k^{s}dk^{2}%
        d^{s-1}\Omega,
    \end{split}
        \label{rate2}
\end{equation}%
where the integral on $k$ includes the term
\begin{equation}
\delta(\omega _{\mathbf{k}}^{\pm }-\omega _{\mathbf{k}_{2}})\simeq \delta \left(\frac{c^{2}}{4\omega_p}((
1+k_{\parallel }^{2}/k^{2})k^{2}-k_{\pm }^{2})\right),
\label{dirac}
\end{equation}%
with
\begin{equation}
k_{\pm }^{2}\lambda_D^2=\frac{2v_T^{2}}{c^{2}}\left( 3k_{2}^{2}\lambda
_{D}^{2}\mp \frac{\omega _{c}}{\omega _{p}}\right)>0.  \label{kpm}
\end{equation}%
Note the important fact that $k_{\pm }^{2}$ has to be positive for electromagnetic waves to be emitted. It means that for $\mathcal{X}$-mode waves, the condition $\omega _{c}/\omega
_{p}<3k_{2}^{2}\lambda _{D}^{2}$ has to be satisfied; if
not, this mode can only be radiated at very low energy. After integration on $k$, we get 

\begin{equation}
    \begin{split}
    &\mu^\pm=\frac{d}{\omega_{p}dt}
    {\displaystyle\int_{V_s}}
    \frac{d^{s}\mathbf{k}}{(2\pi)^{s}}\left\langle \left\vert B_{\mathbf{k}}^{\pm
    }(t)\right\vert ^{2}\right\rangle \simeq\frac{V_{s}}{\lambda_{D}^{s}2^{s/2}
    \pi^{s-1}}\left(  \frac{v_{T}}{c}\right)^{s}\left(  \frac{\omega_{p}^{2}
    }{\omega_{c}^{2}-\omega_{p}^{2}}\right)^{2}\times\\
    &    \times{\displaystyle\int_{V_s}}
    \frac{d^{s}\mathbf{k}_{2}}{(2\pi)^{s}}\left\vert \frac{\delta n_{-\mathbf{k}
    _{2}}\left(  t\right)  }{n_{0}}\right\vert ^{2}\left\vert E_{\mathbf{k}_{2}
    }(t)\right\vert ^{2}\left(  3k_{2}^{2}\lambda_{D}^{2}\mp\frac{\omega_{c}
    }{\omega_{p}}\right)  ^{s/2}\mathcal{I}_{s}(\Omega),    \end{split}
\end{equation}
\label{first rate}
where the double integral $\mathcal{I}_{s}(\Omega)$ on angles (which does not depend on wavevectors' moduli) is given by%
\begin{equation}
    \mathcal{I}_{s}(\Omega)=\int_{\Omega}\left\vert \mathbf{a}_{\mathbf{k}}^{\pm\ast
    }\cdot\eta_{\mathbf{kk}_{2}}^{\prime}\right\vert ^{2}\left(  1+k_{\parallel
    }^{2}/k^{2}\right)  ^{-(s/2+1)}d^{s-1}\Omega.\label{angleint}%
\end{equation}
After analytical integration of (\ref{angleint}) in spherical coordinates ($0\leq \theta \leq \pi ,$ $0\leq \psi \leq 2\pi $, $d^2\Omega=\sin\theta d\theta d\psi$), we get the radiation rates at frequency $\omega _{p}$ in 3D geometry ($s=3$, $V_s=V$)

\begin{equation}
    \begin{split}
    &\mu^\pm=\frac{d}{\omega_{p}dt}
    {\displaystyle\int_{V}}
    \frac{d^{3}\mathbf{k}}{(2\pi)^{3}}\left\langle \left\vert B_{\mathbf{k}}^{\pm
    }(t)\right\vert ^{2}\right\rangle   \simeq\frac{V}{\pi\lambda_{D}^{3}}\left(
    \frac{v_{T}}{c}\right)  ^{3}\left(  \frac{\omega_{p}}{\omega_{p}\mp\omega_{c}%
    }\right)  ^{2}\times\\
    & \times{\displaystyle\int_{V}}
    \frac{d^{3}\mathbf{k}_{2}}{(2\pi)^{3}}\left\vert \frac{\delta n_{-\mathbf{k}%
    _{2}}\left(  t\right)  }{n_{0}}\right\vert ^{2}\left\vert E_{\mathbf{k}_{2}%
    }(t)\right\vert ^{2}\left(  3k_{2}^{2}\lambda_{D}^{2}\mp\frac{\omega_{c}}{\omega_{p}%
    }\right)  ^{3/2}\left(  \sin^{2}\theta_{2}+\frac{\cos^{2}\theta_{2}}%
    {15}\left(  1\pm\frac{\omega_{c}}{\omega_{p}}\right)  ^{-2}\right),
    \end{split}
    \label{ratefinal}
\end{equation}
with $\cos{\theta_2}=\mathbf{h}\cdot{\mathbf{k}_2}/k_2$. Finally, taking into account that $\omega_c \ll \omega_p$, radiation rates appear as a double integral on $\mathcal{LZ}$ wavevectors $\mathbf{k}_2$($k_2$,$\theta_2$,$\psi_2$), which includes the spectral energy $ \left\vert E_{\mathbf{k}_{2}
    }(t)\right\vert ^{2}$ of turbulent wavefields and the plasma parameters (density fluctuations' spectrum $ \left\vert \delta n_{-\mathbf{k}_{2}
    }(t)\right\vert ^{2}$ of average level  $\Delta N$, electron thermal velocity $v_T$, and magnetization $\omega_c/\omega_p$)
\begin{equation}
\begin{split}
    &\mu^\pm=\frac{d}{\omega_{p}dt}
    {\displaystyle\int_V}
    \frac{d^{3}\mathbf{k}}{(2\pi)^{3}}\left\langle \left\vert B_{\mathbf{k}}^{\pm
    }(t)\right\vert ^{2}\right\rangle \simeq\frac{V}{\pi\lambda_{D}^{3}}\left(
    \frac{v_{T}}{c}\right)  ^{3} \times\\
    & \times    {\displaystyle\int_V}
    \frac{d^{3}\mathbf{k}_{2}}{(2\pi)^{3}}\left\vert \frac{\delta n_{-\mathbf{k}
    _{2}}\left(  t\right)  }{n_{0}}\right\vert ^{2}\left\vert E_{\mathbf{k}_{2}
    }(t)\right\vert ^{2}\left(  3k_{2}^{2}\lambda_{D}^{2}\mp\frac{\omega_{c}
    }{\omega_{p}}\right)  ^{3/2}\left(  \sin^{2}\theta_{2}\left(  1\pm
    2\frac{\omega_{c}}{\omega_{p}}\right)  +\frac{\cos^{2}\theta_{2}}{15}\right).
\end{split}
\label{ratefinal2}
\end{equation}

Note that the radiation rates $\mu^\pm$  scale as $(v_T/c)^3$ ($(v_T/c)^2$) in 3D (2D) geometry.

\section{Code availability}
All information on the Particle-In-Cell code SMILEI can be found at 
\href{https://smileipic.github.io/Smilei/}{https://smileipic.github.io/Smilei/}, i.e. the code itself (in C++) 
with a complete description of its internal structure (algorithms and parallelization used), the resources required (modules) to compile it on a large number of supercomputers as well as the initialization method. 

\section{Data availability}
The datasets generated and analyzed in this work are produced via two distinct approaches: (i) 2D/3V full-particle simulations and (ii) solving differential equations. The former ones can be found at the Zenodo repository (\href{https://doi.org/10.5281/zenodo.14900142}{zenodo.14900142}) with python programs for reproducing the corresponding figures. For the latter ones, the code used to generate the datasets is currently in preparation to be made publicly available. In the interim period, the code can be made available from the corresponding author on reasonable request.

\section{Acknowledgements}
 This work was granted access to the HPC computing and storage resources under the allocation 2023-A0130510106 and 2024-A017051010 made by GENCI. This research was also financed in part by the French National Research Agency (ANR) under the project ANR-23-CE30-0049-01.  C.K. thanks the International Space Science Institute (ISSI) in Bern through ISSI International Team project No. 557, Beam-Plasma Interaction in the Solar Wind and the Generation of Type III Radio Bursts. C. K. thanks the Institut Universitaire de France (IUF).
 
\section{Open Access}
For open access purposes, a CC-BY license has been applied by the authors to this document and will be applied to any subsequent version up to the author's manuscript accepted for publication resulting from this submission.

\printbibliography[]

\end{document}